26th International Conference on Science and Technology Indicators | STI 2022

"From Global Indicators to Local Applications"

7-9 September 2022 | Granada, Spain

#STI22GRX

# Measuring web connectivity between research organizations through ROR identifiers[1]

Enrique Orduña-Malea[*] and Núria Bautista-Puig[**]

[*] *enorma@upv.es*
Department of Audiovisual Communication, Documentation and History of Art, Universitat Politècnica de València, Camino de Vera s/n, Valencia, 46022 (Spain)

[**] *nbautist@bib.uc3m.es*
Department of Library and Information Science, Universidad Carlos III de Madrid, Madrid 126, Getafe, 28903 (Spain)

**Introduction**

Digital information needs to be accessed and used in a manageable and sustainable manner to facilitate the advancement of science and science management. Many types of Persistent Identifiers (PIDs) are already in use and well-established in support of the scholarly communication industry, mainly digital objects (e.g. DOIs) and person identifiers (e.g. ORCID). PIDs improve interoperability of digital entities, make them reusable and, at the same time, foster FAIR principles, that is, the capacity of computational systems to Find, Access, Interoperate, and Reuse data with none or minimal human intervention (Cousijn et al., 2021).

PIDs have actually become a priority of the scientific ecosystem, having a specific European policy formulated by the EOSC (European Commission et al., 2020). However, organisational identifiers remained a challenge for a long, having many players attempting to come up with a solution. In this framework, ROR (Research Organization Registry), a community-led collaboration between multiple organisations (Crossref, California Digital Library, DataCite, and Digital Science) that attempts to create identifiers for every research organization, was launched in 2019. This system uses data from Digital Science's GRID database under a CC0 license, having a ROR registry search interface, an Open API and a reconciler that works with OpenRefine[i] covering more than 100,000 research organizations (Lammey, 2020). In addition, ROR currently maps its IDs to other identifiers for the same organization (e.g. GRID, ISNI, Wikidata, and Crossref's Funder ID).

This initiative was born out of a series of collaborative workshops that started back in 2016, during which representatives of different organizations (publishers, libraries, platform providers, metadata services, and other stakeholders) met in different meetings (e.g. FORCE11 conference in Portland, 2016) to define a common vision[ii], and to seek expressions of interest (request for information) from other organizations interested in getting involved with the ROR initiative.

---

[1] This work was supported by the research project Universeo (ref. GV/2021/141), funded by the regional government of the Valencian Community (Consellería Valenciana-GVA).

A set of six governance recommendations and ten product principles were approved by the Organization Identifier Working Group in 2017[iii], and a draft proposal meeting the Steering Group requirements was finally launched, in which the ROR promoters agreed to implement the pilot, feeding the registry with the GRID database, and ensuring a community involvement[iv].

In September 2021, GRID completed its last update, being retired from the public space by Digital Science. ROR incorporated this release into its own data, allowing a 1:1 correspondence of GRID IDs to ROR IDs, and vice versa, and started to be maintained independently as a leading open organization identifier[v]. By March 2022, ROR includes more than 100,000 registries, provides access via an online ROR registry Search, a ROR API, and a ROR data dump. It has been also included in the DataCite, Crossref and ORCID datasets as a disambiguated Organization ID. The ROR journey is summarized in Figure 1.

Figure 1. The journey of ROR.

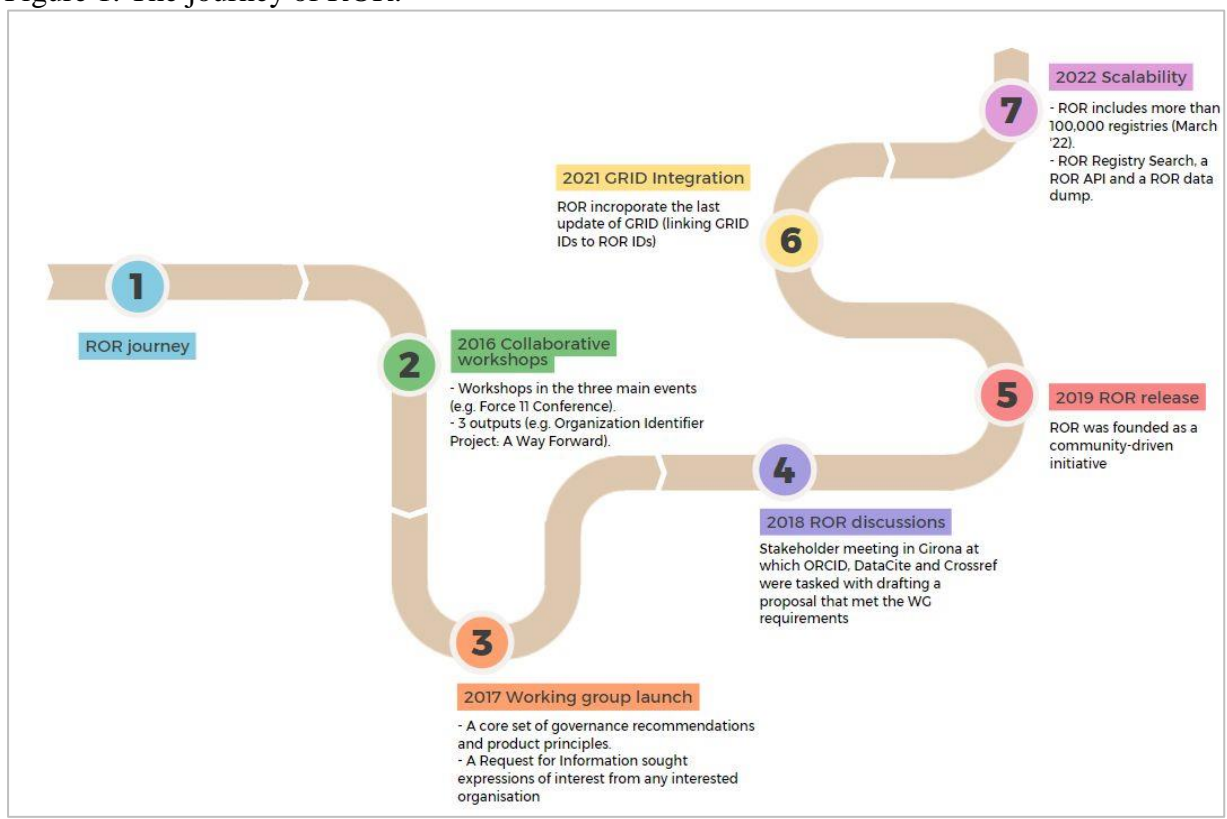

As with ORCIDs and DOIs, the ROR identifier can be embedded in a URL (e.g., https://ror.org/03vek6s52), which we referred to as a ROR-based URL. This ROR-based URL can be linked from any webpage driving users to a customized webpage (ROR card) that includes basic information about the corresponding organization (Figure 2). This URL implementation enables the possibility of carrying out link analysis (Thelwall, 2004), considering each ROR card as a linked webpage, available under a general domain name (i.e., ror.org).

Figure 2: Example of a ROR ID webpage card.

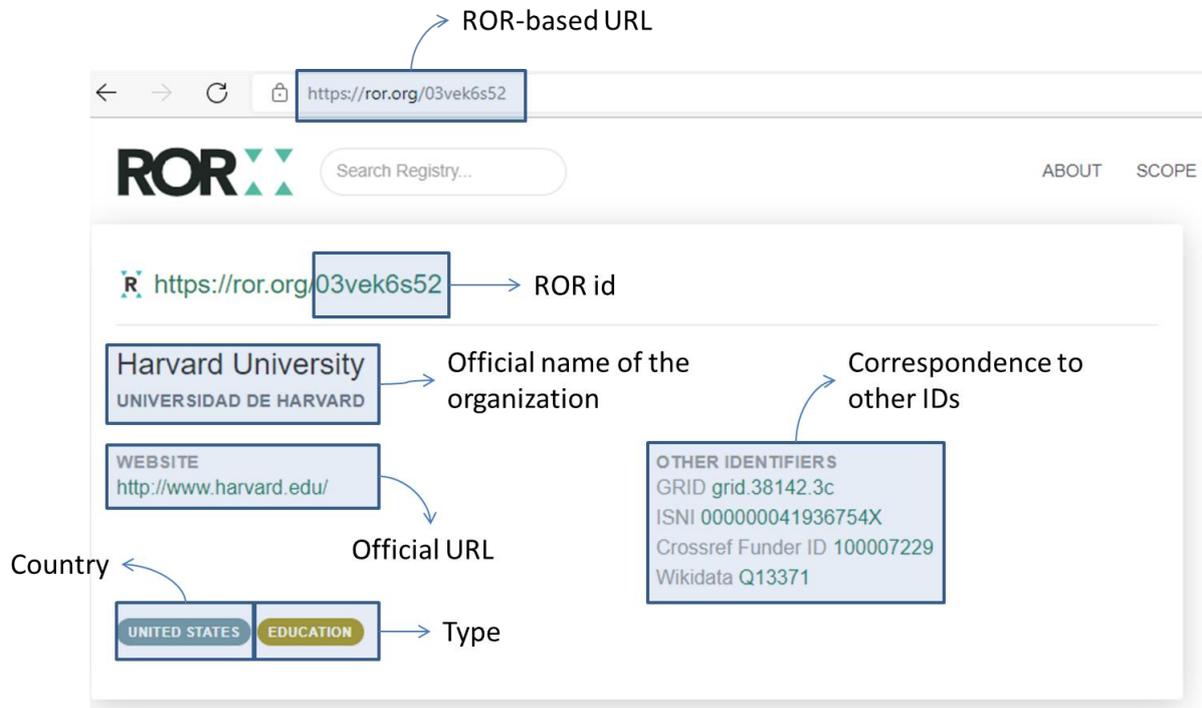

Link-based indicators (e.g., links counts, referring domains counts) to ROR-based URLs might provide information related to the use of ROR identifiers by the scholarly and academic community. The interpretation of these indicators would differ depending on the specific use of each link.

The main objective of this exploratory work is to measure the degree and type of use of ROR identifiers by the online scientific and academic ecosystem through link-based indicators. The specific research goals are set below:

a) How much are ROR-based URLs used? That is, which percentage of ROR ids is linked? Are there differences by country or type of research organization?
b) Who links to ROR-based URLs? That is, which websites (either research or non-research institutions) link to ROR ids?
c) Why ROR-based URLs are linked? That is, which are the main uses for the creation of links to ROR ids?

**Method**

The official data dump feature offered by Research Organization Registry (2022) was used to collect all the metadata related to each ROR identifier. The last available data dump version (March 17, 2022) was downloaded and exported to a spreadsheet for further descriptive statistical analysis. The following metadata fields were analysed: *ROR id* (Unique ROR id for the organization), *name* (the primary name of the organization), *established* (year the organization was established), *country* (country where organization is located), *type* (Education,

Healthcare, Company, Archive, Non-profit, Government, Facility, Other), *link* (official website of the organization), and *status* (whether the organization is active or not).

This process yielded a total of 102,559 ROR identifiers, which constitute the baseline for this study. 98.5% (101,015) of these ROR IDs included a link to the official research organization website. As different research organizations might share the same official website (e.g., child organizations within a mother organization), a total of 93,862 unique domain names related to research organizations were finally found.

The link intelligent professional tool Majestic (https://majestic.com) was used to collect link-based data for each ROR identifier. Due to the dynamism of the Web, the Majestic's fresh database (covering data from December 2021 to March 2022) was used to obtain a current and updated picture of the Web and to minimize the effect of deleted links (i.e., links that existed but have currently been removed) as much as possible. All links received by any webpage published under the official ROR website (*.ror.org/*) were collected as of March 3, 2022. For each link, both the linking page (the source page and the corresponding referring domain) and the linked page (the research organization) were identified and analysed.

A data filtering process was then carried out, which basically consisted in merging http and https versions, and limiting each combination of linking page/linked page to one link. This process yielded a total of 149,851 links to ror.org webpages: 147,154 links to ROR-based URLs, and 2,698 links to other informative webpages under the ror.org website.

Finally, all 147,153 links to ROR-based URLs were categorized according to the link usage. To do this, a bottom-up process was carried out by the two authors. This process consisted of two iterations. At the first iteration, each linking page was manually accessed and categorized according both to the type of the document (e.g., publication, bibliographic record, organization record, etc.) and the link role and location (e.g., author affiliation, organization metadata field, acknowledgement field, etc.), creating thus the main categories. After this iteration, the categories were discussed (merging and splitting if necessary) to find out a final consensual classification scheme. The second iteration consisted of refining the previous categorization.

The categorization process provided the following potential uses to link ROR-based URLs (Table 1):

Table 1. Link categories of ROR-based URLs.

| Category | Scope |
|---|---|
| Acknowledgement in bibliographic record | The acknowledgement/funding section of a bibliographic record includes a ROR-based URL to identify the organization mentioned. |
| Acknowledgement in publication | The acknowledgement/funding section of a publication includes a ROR-based URL to identify the organization mentioned |
| Author affiliation in bibliographic record | A bibliographic record includes a ROR-based URL to identify the institutional affiliation of the authors of the publication |
| Author affiliation in bio field | Online content (e.g., blog post, forum post, wiki) includes a ROR-based URL as part of the biographical field of one author. |
| Author affiliation in other material | Online content (e.g., blog post, forum post, wiki) includes a ROR-based URL to identify the institutional affiliation of the author of the online content. |
| Author affiliation in publication | A scientific publication includes a ROR-based URL to identify the institutional affiliation of the author of the publication. |
| Author affiliation in profile | An author profile includes a ROR-based URL as part of the institutional affiliation data. |
| Self-identification | A webpage includes a ROR-based URL in the footer, header, or sidebar to identify the organization. |
| Example in publication | The body text of a publication (e.g., a table, a paragraph) includes a ROR-based URL to identify the organization mentioned. |
| Example in other contents | The body text of online content (e.g., blog post, forum post, wiki) includes a ROR-based URL to identify the organization mentioned. |
| Organization card | A technical sheet describing an organization includes a ROR-based URL as a metadata field. |
| Organization field in product card | A technical sheet describing a product (e.g., repository) includes a ROR-based as a metadata field. |
| Research assistance affiliation | Individuals supporting academic-related activities (e.g., journal editors, reviews, conference organization) include a ROR-based URL as part of their institutional affiliation data. |
| Support | Online content includes a ROR-based URL to another organization recognizing its support or collaboration. |

**Findings**
*Use of ROR-based URLs*
ROR identifiers have been created for research organizations in 220 countries. However, the distribution of ROR identifiers per country is highly skewed, as only 20 countries represent 80.9% of all ROR identifiers created. The United States (30% of all ROR IDs created and active) and the United Kingdom institutions (7.16%) are the countries with most active ROR IDs. The creation of ROR IDs in Africa, Latin America (except Brazil) and part of Asia is still low (Figure 3).

51.6% of all active ROR identifiers have been linked at least once. This percentage varies remarkably according to each country. Considering those countries with at least 1,000 ROR IDs, 70.6% of Norway's ROR IDs have been linked. On the opposite, the Chinese and Japanese ROR IDs have been scarcely linked (35.8% and 39.6, respectively).

Figure 3: Distribution of ROR identifiers by country.

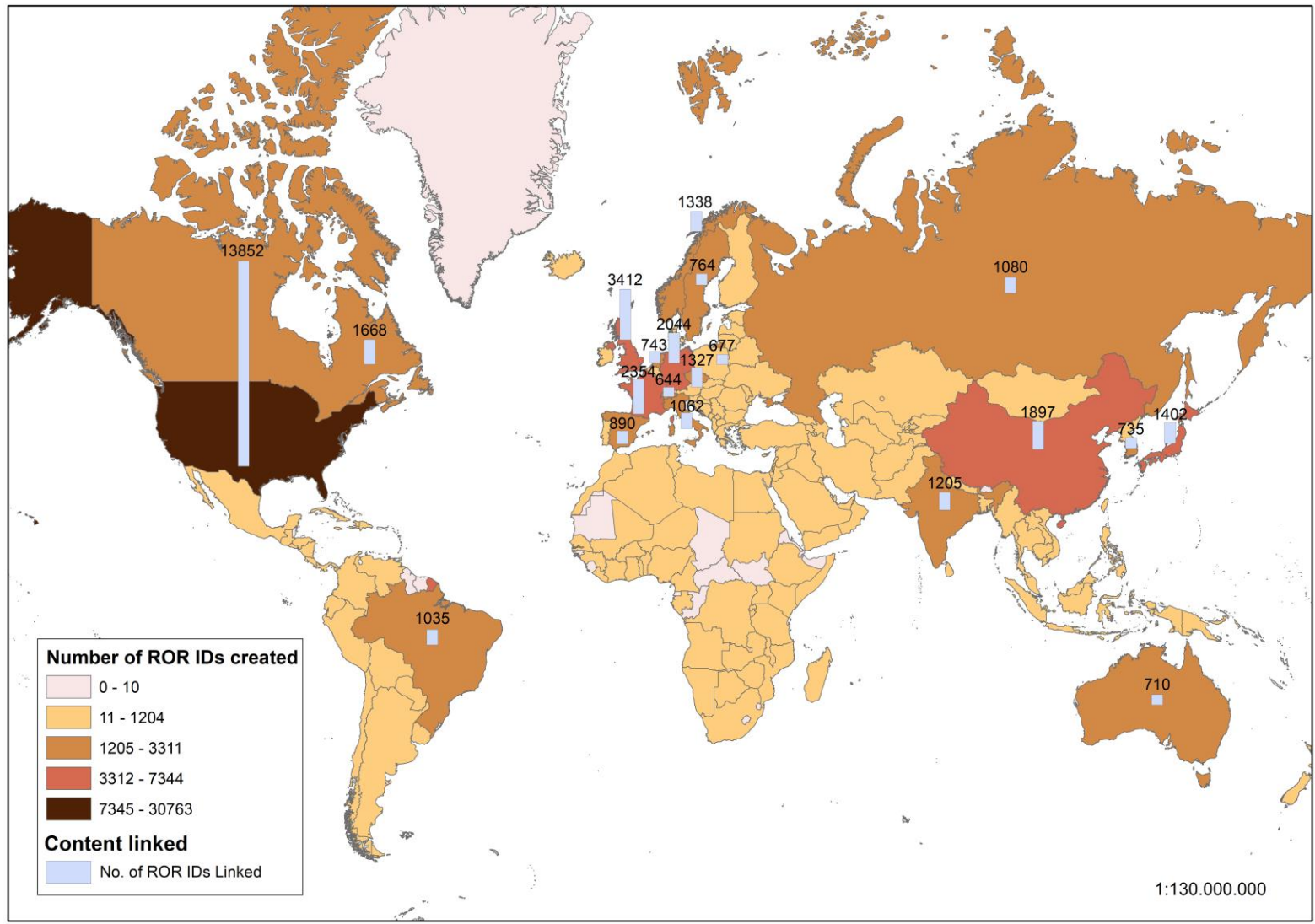

Note: Only countries with >500 Number of ROR IDs linked are displaye

Companies are the most frequent research organization type (28.77%), followed by Education (19.37) and Non-profit organizations (13.47%). The percentage of ROR IDs linked also varies by research organization type (Table 2), obtaining higher values for education organizations (60.1%), and lower for Archives (39%), Non-profit organizations (44.5%) and Companies (48.4%).

Table 2. Types of research organizations and link distribution.

| Type | No. ROR IDs registered | Distribution (%) | No. RORs linked | Linkage (%) | No. Links received | No. Referring domains |
|---|---|---|---|---|---|---|
| Company | 29,508 | 28.77 | 14,287 | 48.4 | 15,477 | 26 |
| Education | 19,866 | 19.37 | 11,930 | 60.1 | 64,407 | 162 |
| Non-profit | 13,817 | 13.47 | 6,146 | 44.5 | 8,888 | 53 |
| Healthcare | 12,978 | 12.65 | 6,500 | 50.1 | 6,831 | 25 |
| Facility | 9,330 | 9.10 | 5,258 | 56.4 | 32,748 | 72 |
| Other | 8,130 | 7.93 | 3,279 | 40.3 | 4,668 | 26 |
| Government | 6,066 | 5.91 | 3,098 | 51.1 | 12,009 | 54 |
| Archive | 2,833 | 2.76 | 1,116 | 39.4 | 2,018 | 46 |
| Blank | 31 | 0.03 | 18 | 58.1 | 18 | 11 |
| **TOTAL** | 102,559 | 100 | 51,632 | 50.3 | 147,064 | *242 |

* Unique referring domains

The number of links received by ROR-based URLs also shows differences between countries and research organization types. On the one hand, the United States is the country that receives most links to their research organizations (42,469), followed by Russia (20,028) and Germany (16,360). On the other hand, educational institutions are the research organizations receiving most links (64,404) from 162 unique referring domains. Companies exhibit a lower visibility, receiving 15,477 links from 26 unique referring domains. However, these aggregated results are biased towards the performance of few specific research organizations (receiving a great quantity of links; see Table 2) and few specific referring domains (linking massively to ROR-based URLs; see Table 3).

The Keldysh Institute of Applied Mathematics, a Russian facility research organization, is the research organization whose ROR-based URL receives most links (9,676). However, these links come from only two domain names, being 'keldysh.ru' (the official website of said institution) one of them, which provides 9,675 links to the corresponding ROR-based URL. Otherwise, the Alfred Wegener Institute for Polar and Marine Research, a German facility research organization, receives 8,103 links from five referring domains, neither of them from its official website. The presence of Russian and German institutions in this ranking (four organizations in the top 20 each) explains the overall presence of these countries mentioned above, and evidence that the most linked ROR-based URLs are linked from few referring domains. On some occasions, links come from the same organization, evidencing self-representation motivation (Table 3).

Table 3. ROR identifiers most linked.

| ROR id | Type | Country | Links | Referring Domains | Self-links |
| --- | --- | --- | --- | --- | --- |
| 01dv3hq14 | Facility | Russia | 9,676 | 2 | 9,675 |
| 032e6b942 | Facility | Germany | 8,103 | 5 | 0 |
| 03tzaeb71 | Education | United States | 7,194 | 5 | 0 |
| 05yc77b46 | Education | Spain | 2,600 | 7 | 2,580 |
| 02z5nhe81 | Government | United States | 2,365 | 6 | 0 |
| 01jmd7f74 | Education | Russia | 1,327 | 3 | 0 |
| 00tgqzw13 | Government | United States | 1,291 | 3 | 0 |
| 001w4ps18 | Education | Russia | 1,254 | 2 | 0 |
| 01qz5mb56 | Facility | United States | 1,021 | 4 | 0 |
| 04ers2y35 | Education | Germany | 949 | 4 | 0 |
| 01gbk3827 | Government | Peru | 814 | 1 | 0 |
| 033n9gh91 | Education | Germany | 798 | 4 | 0 |
| 044jxhp58 | Facility | France | 701 | 4 | 0 |
| 011n96f14 | Facility | Norway | 681 | 2 | 0 |
| 03cjece29 | Education | Russia | 640 | 2 | 0 |
| 009syct46 | Facility | Austria | 578 | 1 | 0 |
| 03zbnzt98 | Non-profit | United States | 577 | 7 | 0 |
| 02772kk97 | Other | Japan | 552 | 3 | 0 |
| 04v76ef78 | Education | Germany | 521 | 6 | 0 |
| 00qps9a02 | Facility | Italy | 404 | 6 | 339 |

*Origin of links to ROR-based URLs*

The 147,154 links targeted to ROR-based URLs come from 242 unique referring domains, representing both research (39.9%) and non-research organizations (60.1%). The weight of non-research organizations can be observed through the remarkable presence of other institutional identifier providers (e.g., grid.ac), journal databases (e.g., sherpa.ac.uk, scielo.org), author databases (e.g., rescognito.com), datasets (e.g., pangaea.de, wikidata.org, suprabank.org) or publishers (e.g., benjamins.com), which evidence the interconnectivity between the ROR identifiers and other scholarly-related online projects (Table 4).

Specifically, the connectivity between ROR and GRID can be observed through the 54,978 links received by ror.org from 54,976 unique webpages from grid.ac. Of these links, 48,435 are targeted to ROR-based URLs–just one link per each ROR ID. Considering that last ROR data dump was still fed from the GRID database, the number of ROR IDs linked is far from the 102,559 ROR IDs registered. Beyond GRID, Pangaea (32,256 links to 227 ROR IDs) and the academic publishing company Benjamins (21,370 links to 2,273 ROR IDs) are the most important referring domains providing links to ROR-based URLs. These three referring domains jointly represent the 69.4% of all links received by ROR-based URLs.

Table 4. Top referring domains linking to ROR identifiers.

| Referring Domain | Category | Inbound links counts | Linking pages count | Number of links to ROR ids | Number of ROR ids linked |
|---|---|---|---|---|---|
| grid.ac | NO | 54,978 | 54,976 | **48,435** | 48,435 |
| pangaea.de | NO | 32,323 | 32,256 | **32,256** | 227 |
| benjamins.com | YES | 21,370 | 21,370 | **21,370** | 2,273 |
| keldysh.ru | YES | 9,685 | 9,685 | **9,680** | 6 |
| esipfed.org | YES | 9,157 | 9,113 | **8,913** | 166 |
| phsreda.com | NO | 8,488 | 8,488 | **8,488** | 86 |
| sherpa.ac.uk | NO | 3,881 | 3,881 | **3,881** | 2,560 |
| uco.es | YES | 2,729 | 2,729 | **2,580** | 1 |
| fdsn.org | NO | 6,039 | 6,039 | **2,452** | 271 |
| knowledia.com | NO | 4,708 | 1,570 | **1,570** | 1,488 |
| wikidata.org | NO | 1,231 | 1,231 | **1,231** | 1,076 |
| ingv.it | YES | 909 | 909 | **909** | 106 |
| ejsei.com | NO | 1,032 | 516 | **516** | 169 |
| tib.eu | YES | 428 | 427 | **390** | 1 |
| journaledu.com | NO | 378 | 378 | **378** | 67 |
| suprabank.org | NO | 328 | 328 | **328** | 99 |
| rescognito.com | NO | 263 | 263 | **263** | 5 |
| scielo.org | YES | 488 | 244 | **244** | 128 |
| ijier.net | NO | 596 | 298 | **235** | 103 |
| docomomojournal.com | NO | 468 | 234 | **234** | 116 |

YES: research organization; NO: non-research organization.

Referring domains exhibit different linking patterns. For example, the University of Córdoba (uco.es) and the Leibniz Information Centre for Science and Technology University Library are massively linking only one ROR ID, precisely their corresponding ROR-based URL (i.e., a self-representation). Other referring domains like GRID or Wikidata provide almost one link per ROR-based URL linked. The different uses behind the creation of links to ROR-based URLs will be detailed in the in the 'Link categories' section.

The flow of links from referring domains to ROR-based URLs can be observed in the Figure 4, where those referring domains providing at least 100 links to ROR-based URLs are displayed. These 26 referring domains provide 145,298 links (98.7% of all links to ROR-based URLs). These results evidence that most of links to ROR-based URLs come from non-research organizations (69.2%; 101,880 links), mainly because of the GRID-ROR connection. The percentage of links from other research organizations is also remarkable (21.6%; 31,777 links), while the self-links (organizations linking to their own ROR-based URL) constitute the smallest percentage of links to ROR-based URLs (9.1%; 13,448 links).

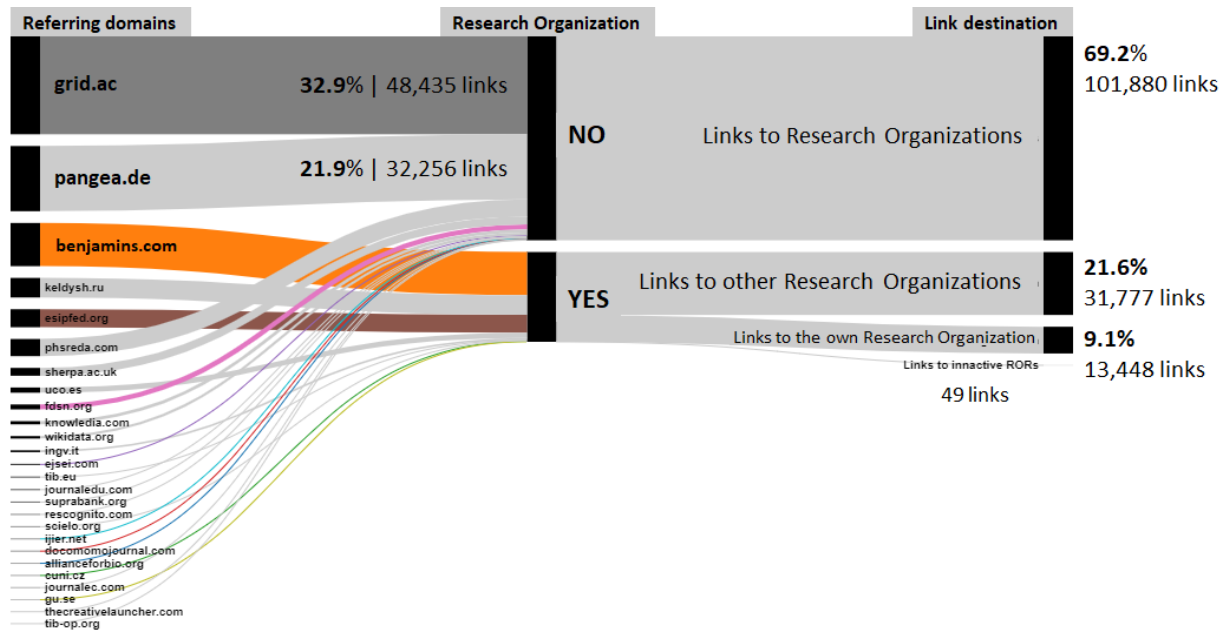

Figure 4: Linking behaviour of referring domains linking to ROR-related webpages.

*Link categories*

The inclusion of ROR-based URLs in bibliographic records (mainly scientific publications and datasets) is the most common use of those links targeted to ROR identifiers (51.4% of all links), followed by the inclusion of ROR-based URLs as part of the identification of the research organization in online institution records (e.g., GRID).

The inclusion of ROR-based URLs as examples in blogs and technical sheets is also remarkable (9,073 links from 57 unique referring domains). Otherwise, it should be mentioned the existence of links not located in the linking page (422) and not available linking webpages (44 links), which evidence the dynamism of the Web (Table 5).

Table 5. Number of links to ROR-based URLs by link category.

| Category | Number of links | % | Number of referring domains |
| --- | --- | --- | --- |
| Author affiliation in bibliographic record | 75,598 | 51.4 | 77 |
| Organization card | 52,928 | 36.0 | 32 |
| Example in other contents | 9,073 | 6.2 | 57 |
| Self-identification | 2,668 | 1.8 | 23 |
| Organization field in product card | 2,538 | 1.7 | 2 |
| Organization field in bibliographic record | 2,452 | 1.7 | 1 |
| Author affiliation in profile | 613 | 0.4 | 4 |
| Support | 463 | 0.3 | 2 |
| ROR not located | 422 | 0.3 | 12 |
| Acknowledgement in bibliographic record | 151 | 0.1 | 3 |
| Research assistance affiliation | 120 | 0.1 | 9 |
| Author affiliation in Bio field | 50 | 0.0 | 3 |
| Not available | 44 | 0.0 | 21 |
| Example in publication | 14 | 0.0 | 5 |
| Acknowledgement in publication | 10 | 0.0 | 3 |
| Author affiliation in other material | 6 | 0.0 | 3 |
| Bibliographic reference | 4 | 0.0 | 1 |
| Total | 147,154 | 100 | 258 |

**Conclusions**

The results obtained in this exploratory work evidence that the percentage of ROR identifiers linked is limited (51.6% of ROR identifiers have been linked at least once). These links come from a limited number of referring domains (242 unique domain names), and mainly from bibliographic records (51.4% of links) and organization cards (36% of links). While the distribution of ROR identifiers is biased towards Anglo-Saxon countries (mainly United States) and type (companies), the educational research organizations are the institutions most linked through their corresponding ROR-based URLs.

This study has covered the number of links towards ROR identifiers considering the whole internet. This approach has allowed to obtain a general overview of the implementation of ROR across the entire online academic ecosystem. Future studies should measure the degree of implementation of ROR IDs on specific bibliographic databases (e.g. DOAJ, Sherpa, Crossref). At the time of writing this work, this implementation is still limited. For example, Crossref has provided coverage data information showing a limited integration of ROR in Crossref Metadata, as of January 2022 (3,790 records include a ROR IDs, covering 205 different ROR IDs)[vi]. The coverage is expected to increase in the following years.

However, the following limitations should be acknowledged. First, the dynamism of the Web causes the existence of deleted links or removed websites. This makes links counts instable and non-cumulative and must always be interpreted as a still picture at the time of data collection. Second, link data depend on the link source. In this case, data have been obtained from Majestic, a professional link intelligent tool.

Results should be restricted to the coverage of this source, as other link sources might provide different results. Third, the use of ROR identifiers can be underrepresented, as they can be inserted without links (e.g., ROR 04q93ds34). Fourth, links counts can be overrepresented (e.g., webpages in different languages might inflate the number of links to specific ROR-based URLs, being the number of referring domains a more accurate metric). These shortcomings must be minimized as much as possible applying severe data cleansing processes.

While links counts to ROR-based URLs are (currently) unrelated to the scientific productivity or impact of organizations, the analysis of links to and from ROR-based URLs have been shown to be useful to map the degree of use and implementation of organization identifiers between institutions and scientific information systems.

Since the creation of ROR is recent (2019), it is estimated that the number of journals, repositories and other databases will gradually incorporate links to ROR-based URLs in their products. The connectivity between DOIs, ORCIDs and RORs can be the spearhead to carry out new webometric studies, of interest to characterize the presence, impact, and interconnection of the global academic Web. Last, their use will increase as long it is embedded and popularised in the papers, as well as when the databases and search engines allow ROR searching as a field tag (similarly to DOI or ORCID searches). This will result in more effective implementation and popularisation on this new identifier, allowing further comprehensive and accurate studies (e.g., DOI-ROR or ORCID-ROR matches).

---

[i] https://scholarlykitchen.sspnet.org/2019/12/04/are-you-ready-to-ror-an-inside-look-at-this-new-organization-identifier-registry/
[ii] https://ror.org/about/
[iii] https://orcid.figshare.com/ndownloader/files/24349103
[iv] https://ror.org/blog/2018-12-02-the-ror-of-the-crowd/
[v] https://www.digital-science.com/press-release/grid-passes-torch-to-ror/
[vi] https://www.crossref.org/blog/a-ror-some-update-to-our-api/